\documentclass[pdftex,twocolumn,epjc3]{svjour3}          % twocolumn

\pdfoutput=1
\pdfoutput=1
\usepackage{color}
\usepackage{amsmath}
 \usepackage{slashed}

\usepackage{comment}
\usepackage[export]{adjustbox}
\usepackage[utf8]{inputenc} %Codificacion utf-8

\newcommand{\be}{\begin{equation}}
\newcommand{\ee}{\end{equation}}
\def\bsp#1\esp{\begin{split}#1\end{split}}

\newcommand{\bea}{\begin{eqnarray}}  
\newcommand{\eea}{\end{eqnarray}}

\RequirePackage[T1]{fontenc}

\smartqed  % flush right qed marks, e.g. at end of proof

\RequirePackage{graphicx}
\RequirePackage{mathptmx}      % use Times fonts if available on your TeX system
\RequirePackage{flushend}
\RequirePackage[numbers,sort&compress]{natbib}
\RequirePackage[colorlinks,citecolor=blue,urlcolor=blue,linkcolor=blue]{hyperref}

\journalname{Eur. Phys. J. C}

\usepackage{amsmath}
\usepackage{amssymb}
\usepackage{tensor}

\begin{document}

\title{Constructing spherically symmetric Einstein-Dirac systems with multiple spinors: Ansatz, wormholes and other analytical solutions}
%\subtitle{}


\title{Spin-dependence of Gravity-mediated Dark Matter in Warped Extra-Dimensions}

\author{Miguel G. Folgado,\thanksref{e1,addr1}
        Andrea Donini \thanksref{e2,addr1}
        \and
        Nuria Rius\thanksref{e3,addr1}
}

\thankstext{e1}{e-mail: migarfol@ific.uv.es}
\thankstext{e2}{e-mail: donini@ific.uv.es}
\thankstext{e3}{e-mail: nuria.rius@ific.uv.es}

\institute{Departamento de F\'isica Te\'orica and IFIC, Universidad de Valencia-CSIC,
C/ Catedr\'atico Jos\'e Beltr\'an, 2, E-46980 Paterna, Spain\label{addr1}
}

\date{Received: date / Accepted: date}

\maketitle

\begin{abstract}
We study the possibility that Dark Matter (DM) particles of spin 0, 1/2 or 1 may interact gravitationally with Standard Model (SM) particles within the framework of a warped Randall-Sundrum (RS) model.
Both the Dark Matter and the Standard Model particles are assumed to be confined to the Infra-Red (IR) brane and only interchange Kaluza-Klein excitations of the graviton and the radion (adopting the
Goldberger-Wise mechanism to stabilize the size of the extra-dimension). We analyze the different DM annihilation channels and find that the presently observed Dark Matter relic abundance, $\Omega_{\rm DM}$, 
can be obtained within the freeze-out mechanism for DM particles of all considered spins.  This extends our first work concerning scalar DM in RS scenarios \cite{Folgado:2019sgz}
and put it on equal footing with our second work in which we studied DM particles of spin 0, 1/2 and 1 in the framework of the Clockwork/Linear Dilaton (CW/LD) model \cite{Folgado:2019gie}.
We study the region of the model parameter space for which $\Omega_{\rm DM}$ is achieved and compare it with the different experimental and theoretical bounds. 
We find that, for DM particles mass $m_{\rm DM} \in [1,15]$ TeV, most of the parameter space is excluded by the current constraints or will be excluded by the LHC Run III or by the LHC upgrade, the HL-LHC. 
The observed DM relic abundance can still be achieved for DM masses $m_{\rm DM} \in [4,15]$ TeV and $m_{G_1} < 10$ TeV for scalar and vector boson Dark Matter. On the other hand, for spin 1/2 fermion Dark Matter, only a tiny region with $m_{\rm DM } \in [4, 15]$ TeV, $m_{G_1}  \in [5,10]$ TeV and $\Lambda > m_{G_1}$ is compatible with theoretical and experimental bounds.  We have also studied the impact of the radion in the phenomenology, finding that it does not modify significantly the allowed region for DM particles of any spin (differently from the CW/LD case, where its impact was quite significant in the case of scalar DM). 
We, eventually, briefly compare results in RS with those obtained in the CW/LD model.
\end{abstract}

%%%%%%%%%%%%%%%%%%%%%%%%%%%%%%%%%%%%%%%%%%%%%%%%%%%%%%%%%%%%%%%%%%%%%%%%%%%
%%%%%%%%%%%%%%%%%%%%%%%%%%%%%%%%%%%%%%%%%%%%%%%%%%%%%%%%%%%%%%%%%%%%%%%%%%%
\section{Introduction}
\label{sec:intro}
%%%%%%%%%%%%%%%%%%%%%%%%%%%%%%%%%%%%%%%%%%%%%%%%%%%%%%%%%%%%%%%%%%%%%%%%%%%

The Standard Model of Fundamental Interactions is a very powerful tool to understand electromagnetic, weak and strong interactions at least up to the energy
scale tested at the LHC. After the discovery of the Higgs boson in 2012 \cite{Aad:2012tfa} the model is complete and it may well be possible that
a huge energy desert above the TeV scale should be crossed before finding some new phenomena. Accelerators much larger than the LHC \cite{Abada:2019lih}
are currently under study in order to explore the energy landscape above the TeV. However, a reasonable hope can drive us in the future: the Standard Model 
on its own is incapable of explaining the observed baryon asymmetry in the Universe; it does not provide a unique mechanism to generate neutrino masses; 
and, more compellingly, it offers no clues at all to what Dark Matter and Dark Energy are. 
Astrophysical and cosmological data (see, e.g., Ref.~\cite{Bertone:2004pz} and refs. therein) point out that a significant amount of the energy density of the Universe
takes the form of non-baryonic matter, i.e. matter with no apparent interaction with the Standard Model matter we are made of but gravity. At present,
we are far from having a clear suspect to fill the r\^ole of a DM particle, though. For this reason, any meaningful extension of the Standard Model usually includes some DM candidate,
a stable (or long-lived, with a lifetime as long as the age of the Universe) particle with very small or none interaction with Standard Model particles.
These states are usually supposed to be heavy and are called ``WIMP's", or ``weakly interacting massive particles", as it the case of
neutralinos in supersymmetric extensions of the SM \cite{Dimopoulos:1981zb} or the lightest Kaluza-Klein particle in Universal Extra-Dimensions \cite{Appelquist:2000nn}. The typical range of masses for these particles was expected to be $m_{\rm DM} \in [100,1000]$ GeV. However, LHC searches for heavy particles constrain significantly the masses of the candidates, pushing them into the multi-TeV region. 
Experiments searching for DM particles through their interactions
with a fixed target, known as ``Direct Detection" (DD) experiments (see, e.g., Ref.~\cite{Cushman:2013zza}), or through their annihilation into Standard Model particles, known as
``Indirect Detection" (ID) experiments (see, e.g., Ref.~\cite{Cirelli:2010xx}), have thoroughly explored the 
$m_{\rm DM} \in [100,1000]$ GeV region, pushing constraints on the interaction cross-section between DM and SM particles to very 
small values. Notice that both DD and ID experiments have a limited sensitivity above the TeV, as they have been mostly designed to look for ${\cal O} (100)$ GeV particles.  
For all of this, it seems interesting to explore further the possibility that DM is indeed made of WIMPy-like particles with masses in the multi-TeV range and none or 
very small interaction with SM particles beside for their gravitational interaction. 

Four-dimensional gravitational interaction is, however, too weak to explain the observed DM abundance in the Universe for multi-TeV particles. A way out to this 
problem is to enhance the gravitational interaction by lowering the fundamental scale of gravity. This is easily done in any extra-dimensional setup: 
if gravity feels more than 4 dimensions, than the Planck mass $M_{\rm P}$ is only an effective scale relevant for processes at too large distances (or too small energies) to 
test the fundamental scale $M_{\rm D}$. Several extra-dimensional models have been proposed in the last twenty years to solve the "Hierarchy Problem", i.e. the large hierarchy between the electro-weak scale, $\Lambda_{\rm EW} \sim 250$ GeV, and the Planck scale, $M_{\rm P} \sim 10^{19}$ GeV.
Extra-dimensional models solve the hierarchy problem by either replacing the Planck scale $M_{\rm P}$ with a fundamental 
gravitational scale $M_{\rm D}$ (being $D = 4 + n$ the number of dimensions and $n$ the number of extra spatial dimensions) 
that could be as low as a few TeV (Large Extra-Dimensions models, or LED, see Refs.~\cite{Antoniadis:1990ew,Antoniadis:1997zg,ArkaniHamed:1998rs,Antoniadis:1998ig,ArkaniHamed:1998nn}), or by "warping" the space-time such that the effective Planck scale $\Lambda$ felt by particles of the SM is indeed much smaller than the fundamental scale $M_{\rm D} \sim M_{\rm P}$ (see Refs.~\cite{Randall:1999ee,Randall:1999vf}), or by a mixture of the two options (see Refs.~\cite{Giudice:2016yja,Giudice:2017fmj}). 
Gravitational {\em enhancement} of Dark Matter interaction with SM particles was first studied in the framework of RS models (see Refs.~\cite{Lee:2013bua,Lee:2014caa} 
and  Refs.~\cite{Han:2015cty,Rueter:2017nbk,Rizzo:2018obe,Rizzo:2018joy,Carrillo-Monteverde:2018phy,Goyal:2019vsw}). The generic conclusion of these papers was
that, when all the matter content is localized in the so-called TeV (or infrared brane), after taking into account current LHC bounds
it was not possible to achieve the observed Dark Matter relic abundance in warped models for scalar DM particles (whereas this was not the
case for fermion and vector Dark Matter). However, an important caveat was that these conclusions were drawn assuming the DM particle being {\it lighter} 
than the first Kaluza-Klein graviton mode. In this case, the only kinematically available channel to deplete the Dark Matter density in the Early Universe 
is the annihilation of two DM particles into two SM particles through virtual KK-graviton exchange. 
In Ref.~\cite{Folgado:2019sgz}, we studied the particular case of scalar DM in warped extra-dimensions allowing for DM particles to be  {\it heavier} than the first KK-graviton mode.
In this case, annihilation of two DM particles into two KK-gravitons becomes kinematically possible and, through this channel, the observed relic abundance can indeed be achieved 
in a significant region of the parameter space within the freeze-out scenario. Radion exchange and DM annihilation into radions (added as in the Goldberger-Wise mechanism \cite{Goldberger:1999uk}
to stabilize the size of the extra-dimension) were also taken into account, showing in which part of the parameter space they may contribute or not to achieve the relic abundance.
Recent papers studying different aspects of spin-2 mediation of the interaction between DM particles and the
Standard Model  have been published in Refs.~\cite{Kang:2020huh,Chivukula:2020hvi,Kang:2020afi}.

A similar analysis was carried on in Ref.~\cite{Folgado:2019gie} in the framework of the CW/LD model. Also there it was shown that DM (represented by either 
scalar, fermion or vector boson particles) on the IR-brane coupled gravitationally with the SM may achieve the observed relic abundance through the freeze-out mechanism. 
In order to put on equal footing the Randall-Sundrum and the Clockwork/Linear Dilaton models, we extend in the present paper our work of Ref.~\cite{Folgado:2019sgz} (where only
the scalar DM case was studied) to the case in which DM particles can be either scalar, spin 1/2 fermions or vector bosons.
The region of the parameter space for which the observed DM relic abundance is achieved in the freeze-out framework for scalar and vector boson DM particles 
corresponds to  DM masses in the range $m_{\rm DM} \in [1, 15]$ TeV, with the first KK-graviton mass ranging from hundreds of GeV to tens of TeV.  On the other hand, 
we found that it is very difficult to achieve the observed relic abundance for spin 1/2 fermion DM (only a tiny region of the parameter space with $m_{\rm DM} \sim m_{G_1} \sim$ a few TeV
and $\Lambda \sim 1$ TeV survives after taking into account the LHC Run III bounds). In most part of the allowed parameter space, however, the effective gravitational scale $\Lambda$ 
for which interactions between SM particles and KK-gravitons occur must be larger than 10 TeV, approximately. 
Therefore, in this scenario, the hierarchy problem cannot be completely solved and some hierarchy between $\Lambda$ and $\Lambda_{\rm EW}$ is still present. 
This is something, however, common to most proposals of new physics aiming at solving the hierarchy problem, as the LHC has found no hint whatsoever of new physics to date. 
As it was the case in our previous analysis for scalar DM in warped extra-dimensions, a large part of the allowed parameter space (almost all of it, in the case of spin 1/2 fermion DM) 
will be tested using the LHC Run III and the HL-LHC data. By the end of the next decade, therefore, the possibility that DM is indeed made of WIMPy particles
that interact only gravitationally in an extra-dimensional framework can be fully explored.

Notice that a different approach to gravitational coupling of DM to the SM was followed in the recent Ref.~\cite{Bernal:2020fvw}, where it was studied the possibility that scalar 
DM in a Randall-Sundrum scenario is only feebly interacting with the SM and, thus, it never reaches thermal equilibrium. It was shown that the observed relic abundance may be achieved 
also in this case through the so-called {\em freeze-in mechanism} (more details can be found in Ref.~\cite{Hall:2009bx}). These results were then extended to the case
of scalar DM in a CW/LD framework, finding similar results \cite{Bernal:2020yqg}.

The paper is organized as follows: 
%in Sect.~\ref{sec:CW} we outline the theoretical framework, reminding shortly the basic ingredients of the 
%ClockWork/Linear Dilaton extra-dimensional scenario and of how dark matter can be included within this hypothesis; 
in Sect.~\ref{sec:annihilres} we show our results for the annihilation cross-sections
of DM particles into SM particles, KK-gravitons and radion/KK-dilatons;  
in the first part of Sect.~\ref{sec:results} we review the present experimental bounds on the parameters of the model (the effective Planck scale $\Lambda$, the mass of the first KK-graviton, $m_{G_1}$
and the DM mass $m_{\rm DM}$) from the LHC and from direct and indirect searches of Dark Matter, and recall the theoretical constraints 
(coming from unitarity violation and effective field theory consistency); 
in the second part of Sect.~\ref{sec:results} we explore the allowed parameter space such that 
the correct relic abundance is achieved for DM particles; 
and, eventually, in Sect.~\ref{sec:con} we conclude. 
%In the Appendices we give some of the mathematical expressions used in the paper: 
%in App.~\ref{app:spin2} we give the expression relative to the KK-graviton propagator and polarization tensor; 
In App.~\ref{app:feynman} we give the Feynman rules for the theory considered here.
%in App.~\ref{app:decay} we give the expressions for the decay amplitudes of the KK-graviton; 
%in App.~\ref{app:kksum} we remind how the sum over KK-modes is carried on.
%and, eventually, in App.~\ref{app:annihil} we give the formul\ae \, relative to the annihilation
%cross-sections of Dark Matter particles into Standard Model particles, KK-gravitons and radion/KK-dilatons.
Complete expressions for KK-gravitons and radion decay amplitudes and DM annihilation cross-sections into SM particles, KK-gravitons and/or radions in the small relative velocity approximation 
can be found in Ref.~\cite{Folgado:2019gie} and will not be repeated here.

\section{DM annihilation cross-section in RS model}
\label{sec:annihilres}

Experimental data from a wide range of length scales clearly show that a non-negligible component of the Universe energy density is represented by some form of matter
that do not interact electromagnetically (conventionally called {\em non-baryonic}, in cosmologists jargon). 
This component  is called {\em Dark Matter} and, in the cosmological $\Lambda$CDM \cite{Akrami:2019izv} ``standard model",  
is usually assumed to consist of stable (or long-lived) heavy particles, {\em i.e.} non-relativistic (or ``Cold") Dark Matter. 
The Standard Model matter and the Dark Matter component are considered in thermal equilbrium within the freeze-out scenario
(differently from the case of the freeze-in scenario, in which the DM has never been in equilibrium with the Standard Model). 
The evolution of the Dark Matter density $n_{\rm DM}$ follows the following Boltzmann equation \cite{Kolb:1990vq}:
\be
\frac{dn_{\rm DM}}{dt} = -3 H(T) \, n_{\rm DM} - \left\langle \sigma {\rm v} \right\rangle \left [ n_{\rm DM}^2 - (n_{\rm DM}^{eq})^2 \right] \, , 
\label{boltzmann_equation}
\ee
where $T$ is the temperature,  $H(T)$ is the Hubble parameter as a function of the temperature, 
and $n_{\rm DM}^{eq}$ is the DM number density at equilibrium (see Ref.~\cite{Kolb:1990vq} for an explicit expression for $n_{\rm DM}^{eq}$). 

Eq.~(\ref{boltzmann_equation}) depends on two factors: the first proportional to the Hubble expansion rate at temperature $T$, and the second
to the thermally-averaged cross-section, $\left\langle \sigma {\rm v} \right\rangle$. During the expansion of the Universe, 
the thermally-averaged annihilation cross-section  times the number density falls below the Hubble expansion rate,  $\left\langle \sigma {\rm v} \right\rangle \times n_{\rm DM}^2 < H(T)$,
and $n_{\rm DM} (T)$ freezes out.
At that moment, the DM decouples from the SM particles bath and its density in the co-moving frame {\em freezes} to a constant density called DM relic abundance.
The experimental value of the relic abundance in the $\Lambda$CDM model is $\Omega_{\rm CDM} h^2 = 0.1198 \pm 0.0012$, $h$ being the present value of the Hubble parameter (see Ref.~\cite{Aghanim:2018eyx}). 
Solving eq.~(\ref{boltzmann_equation}) we may find, then, the thermally-averaged cross-section at  freeze-out\footnote{
Notice that $\Omega_{\rm DM}$ does not depend on the value of the DM mass  for $m_{\rm DM} >  10$ GeV, and, therefore, 
the value of  $\sigma_{\rm FO}$ needed to obtain the correct relic abundance is insensitive to $m_{\rm DM}$.
} 
$\left\langle \sigma_{\rm FO} \, {\rm v} \right\rangle = 2.2 \times 10^{-26}$ cm$^3$/s \cite{Steigman:2012nb}.
 
 In order to obtain this quantity, we first compute the total annihilation cross-section of the DM particles:
\begin{eqnarray}
\label{eq:sigmaDMtot}
\sigma_{\rm th} &=& \sum_{\rm SM} \sigma_{\rm ve}({\rm DM} \, {\rm DM}  \rightarrow {\rm SM} \, {\rm SM}) + \sigma_{rr} ({\rm DM} \, {\rm DM} \rightarrow r \, r)  \nonumber \\
&+& \sum_{n=1}^\infty \sigma_{Gr} ({\rm DM} \, {\rm DM} \rightarrow G_n \, r) \nonumber \\
 &+& \sum_{n=1}^\infty \sum_{m=1}^\infty \sigma_{GG} ({\rm DM} \, {\rm DM} \rightarrow G_{n} \, G_{m}) \, ,
\end{eqnarray}
where in the first term, $\sigma_{\rm ve}$, the DM particles annihilate through virtual exchange (thus the subscript ${\rm ve}$) through KK-graviton, radion or the Higgs boson\footnote{
The last option is known as  "the Higgs portal" and has been extensively studied in the literature. 
These scenarios are strongly constrained (see for instance \cite{Escudero:2016gzx,Casas:2017jjg}  for recent analyses), 
so we will neglect those couplings and focus only on the gravitational mediators that have not been previously considered.}. 
In this cross-section we sum over all SM particles in the final state and in the KK-graviton modes tower when needed. 
We computed the analytical value of $\left\langle \sigma {\rm v} \right\rangle$ 
using the exact expression from Ref.~\cite{Gondolo:1990dk}:
\be
\left\langle \sigma {\rm v}_{M\o l} \right\rangle =\frac{1}{8m_S^4TK_2^2(x)} \int_{4m_S^2}^{\infty} ds (s-4m_S^2)\, \sqrt{s} \, \sigma(s) \, K_1\left(\frac{\sqrt{s}}{T}\right)\, ,
\label{thermal_average}
\ee
where $K_1$ and $K_2$ are the modified Bessel functions and $ v_{M\o l}$ is the M\o ller velocity.

The second term, $\sigma_{rr}$, corresponds to DM annihilation into radions.
The third term, $\sigma_{Gr}$, corresponds to DM annihilation into one radion and one KK-graviton $G_n$.
Eventually, the fourth term, $\sigma_{GG}$, corresponds to DM annihilation into a pair of KK-gravitons $G_n$ and $G_m$. 

If the DM mass $m_{\rm DM}$ is smaller than the mass of the first KK-graviton $G_1$ and of the radion, only the first channel is possible. 
After that, depending on the mass of the radion with respect to $G_1$, the other channels open. For a radion mass smaller than $m_{G_1}$ (as is usually the case in phenomenological models using the Goldberger-Wise mechanism to stabilize
the size of the extra-dimension), we will take into account in sequence the second, the third and, eventually, the fourth term in eq.~(\ref{eq:sigmaDMtot}). 

A common approximation in the freeze-out paradigm is to consider a small relative velocity v between the DM particles when the freeze-out occurs. Therefore, 
the c.o.m. energy $s$ is usually replace by $s \sim 4 m_{\rm DM}^2$ and only leading order terms in v are kept. 
Formul\ae \, for the DM annihilation into SM particles  in the so-called {\em velocity expansion} were given in Ref.~\cite{Folgado:2019gie} and will be not repeated here. 
We address the interested reader to that reference. 
A final comment is in order:  for mediator masses much smaller than the DM particle mass, annihilation cross-sections into SM particles and/or into the light mediators may be enhanced
by multiple interchange of the light mediator in the initial state, in a phenomenon known as {\em Sommerfeld enhancement} (see, {\em e.g.}, Ref.~\cite{Feng:2010zp} and refs. therein). 
We have not studied in detail the phenomenological impact of this effect, that could be relevant for relatively small radion masses as it may lower the value of $\Lambda$ required to achieve
the observed relic abundance. However, we have estimated that, for radion masses above 100 GeV, the effect should be subdominant in the range of relative velocities v considered here.

\begin{figure*}[htbp]
\centering
\includegraphics[width=140mm]{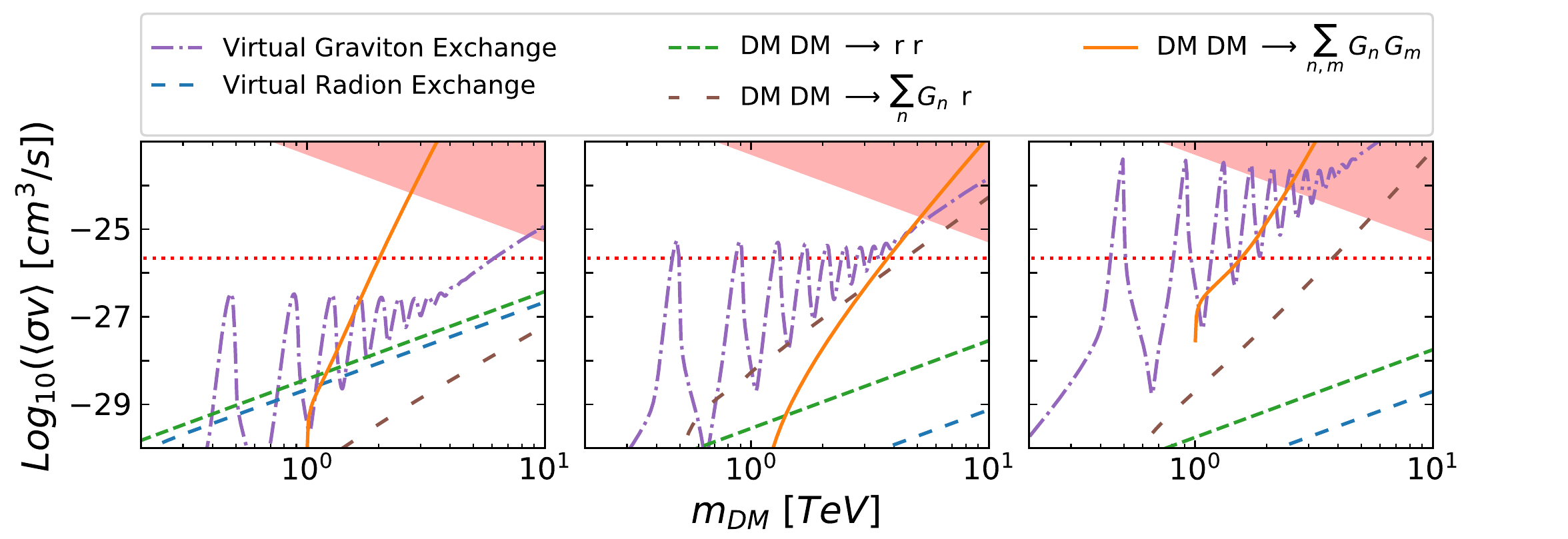}
\caption{\it 
Different contributions to the thermally-averaged annihilation cross-section. The three panels represent (from left to right): 
scalar, spin 1/2 fermion and vector boson DM particles. 
In all cases we consider $m_r = 100$ {\rm GeV}, $m_{G_1} = 1$ {\rm TeV} and $\Lambda = 10$ {\rm TeV}.}
 \label{fig:contribuciones}
\end{figure*}

In Fig.~\ref{fig:contribuciones} we present the different contributions from $\sigma_{\rm ve}$, $\sigma_{rr}$, $\sigma_{Gr}$ and $\sigma_{GG}$ to the thermally-averaged DM annihilation cross-section 
as a function of the DM mass $m_{\rm DM}$  for scalar (left panel), spin 1/2 fermion (middle panel) and vector boson (right panel) Dark Matter particles, respectively. 
The parameters for which the Figure has been obtained are $m_r = 100$ GeV, $m_{G_1} = 1$ TeV and $\Lambda = 10$ TeV. These values have been chosen so as to give a general 
feeling of the typical results that can be obtained. In all plots, 
the freeze-out thermally-averaged cross-section $\left\langle \sigma_{\rm FO} \, {\rm v} \right\rangle $ is depicted by a dotted horizontal (red) line.
The virtual KK-graviton exchange is represented by (purple) dot-dashed lines, and it shows the characteristic spaced multiple-resonances behaviour of the warped scenarios 
(differently from the case of CW/LD model \cite{Folgado:2019gie}, where the spacing  between one KK-graviton mode and the next one is rather small, and a huge number of KK-modes must be coherently summed). We can see in the left panel that, as it was already found in Refs.~\cite{Han:2015cty,Rueter:2017nbk,Rizzo:2018obe,Rizzo:2018joy,Carrillo-Monteverde:2018phy},
for scalar DM the virtual exchange channel is insufficient to reach $\left\langle \sigma_{\rm FO} \, {\rm v} \right\rangle $. This is not the case for fermion and vector boson DM, 
for which the resonant channel dominates the cross-section for DM masses between 1 and 10 TeV. The direct production of two radions, depicted
by a dashed (green) line, is relevant for $m_{\rm DM}$ below 1 TeV in the case of scalar DM, whereas it is much smaller than the resonant channel for fermion and vector bosons. The same
happens for the virtual radion exchange cross-section, depicted by a dashed (blue) line, mostly irrelevant\footnote{
Notice that we have chosen a very small value of $m_r$ so as to study the behaviour of the cross-section outside of the resonant window for the radion mass. 
For radion masses in the range of the DM masses studied here, a resonant peak in the cross-section is obviously found. However, the width of the radion peak 
is so small that a significant fine-tuning should occur in order for $m_{\rm DM} \sim m_r$. We have decided not to consider this particular case in the absence of a theoretical motivation for this fine-tuning relating the mass of the Dark Matter and the mass of the radion.
} in all cases. This is not the case for the direct production of 
one KK-graviton and one radion (represented by a dashed brown line), kinematically possible for $m_{\rm DM} \geq 1/2 m_{G_1}$. In the scalar case this channel is strongly suppressed. 
For vector bosons, $\sigma_{Gr}$ is much smaller than the virtual KK-graviton exchange but much larger than $\sigma_{rr}$ and the virtual radion exchange.
On the other hand, in the fermion case, this cross-section is in the same ballpark of the virtual KK-graviton exchange one and may play a role for $m_{\rm DM} < 1$ TeV. 
The last contribution, depicted by a solid (orange) line, represents the contribution of direct production of two KK-gravitons, kinematically allowed for $m_{\rm DM} \geq m_{G_1}$
(for larger values of $m_{\rm DM}$, new channels open as long as $2 m_{\rm DM} \geq m_{G_m} + m_{G_n}$). For scalar DM, this channel is the driving force to achieve 
$\left\langle \sigma_{\rm FO} \, {\rm v} \right \rangle$ for $m_{\rm DM} > 1$ TeV, as it was found in Ref.~\cite{Folgado:2019sgz}. On the other hand, both for fermion and vector DM, 
this channel 
is of the same order of the virtual KK-graviton exchange and contributes to the total cross-section but is not changing the general behaviour of the latter.
Eventually, the red-shaded area in the upper-right corner represents the region of the parameter space for which the effective field theory we are using here is no longer valid, 
as the cross-section is trespassing the unitarity bound $\left\langle \sigma \, {\rm v} \right\rangle \geq 1/s$.

\begin{table}[]
\begin{center}
\begin{tabular}{c|c|c|c|}
\cline{2-4}
\multicolumn{1}{l|}{}                                      & \multicolumn{1}{l|}{Scalar} & \multicolumn{1}{l|}{Fermion} & \multicolumn{1}{l|}{Vector} \\ \hline
\multicolumn{1}{|c|}{Graviton Virtual Exchange}                 & v$^4$ (d)                          & v$^2$ (p)                              & v$^0$ (s)                              \\ \hline
\multicolumn{1}{|c|}{Radion Virtual Exchange}     & v$^0$ (s)                          & v$^2$ (p)                              & v$^0$ (s)                              \\ \hline
\multicolumn{1}{|c|}{Annihilation into Gravitons}                 & v$^0$ (s)                          & v$^0$ (s)                              & v$^0$ (s)                              \\ \hline
\multicolumn{1}{|c|}{Annihilation into Radions}       & v$^0$ (s)                          & v$^2$ (p)                              & v$^0$ (s)                              \\ \hline
\multicolumn{1}{|c|}{Annihilation into Radion + Graviton}   & v$^0$ (s)                          & v$^0$ (s)                               & v$^0$ (s)                              \\ \hline
\end{tabular}
\end{center}
\caption{\it Velocity dependence of the different DM annihilation channels and the corresponding $s$-, $p$- or $d$-waves.}
\label{tab:orbitals}
\end{table}

As a useful tool  to understand the difference between the cross-sections for scalar, fermion and vector DM particles, we remind in Tab.~\ref{tab:orbitals} the
dependence of the thermally-averaged annihilation cross-section $\langle \sigma \, {\rm v} \rangle$ on the relative velocity v (see Ref.~\cite{Folgado:2019gie}). 
Recall that v acts as a suppression factor and, therefore, the larger the power to which it appears, the smaller the cross-section.

\begin{figure*}[htbp]
\centering
\includegraphics[width=140mm]{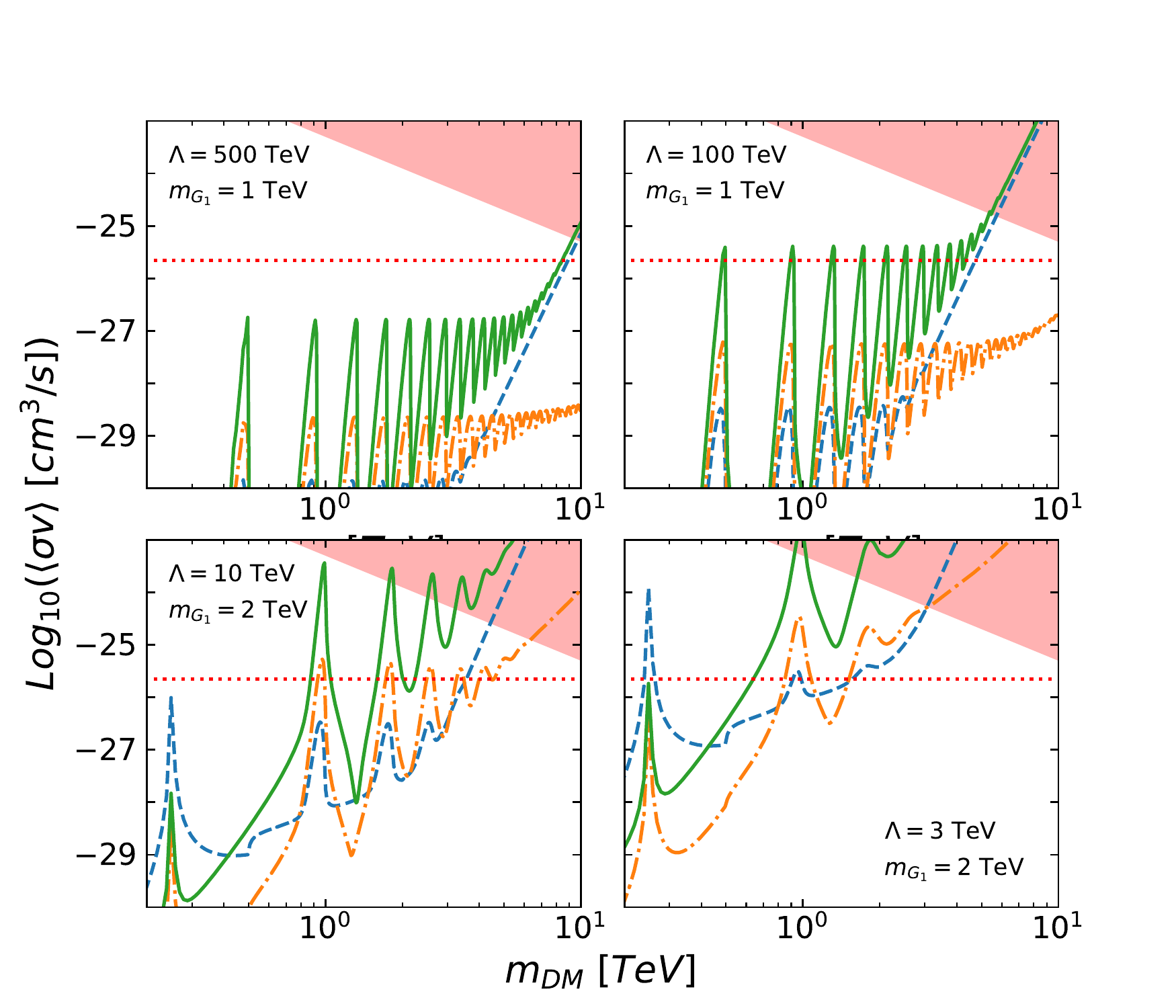}
\caption{\it Several  examples for the total thermally-averaged cross-section as a function of the DM mass $m_{\rm DM}$. 
Upper left panel: $m_{G_1} = 1$ {\rm TeV} and $\Lambda = 500$ {\rm TeV};
Upper right panel: $m_{G_1} = 1$ {\rm TeV} and $\Lambda = 100$ {\rm TeV};
Lower left panel: $m_{G_1} = 2$ {\rm TeV} and $\Lambda = 10$ {\rm TeV};
Lower left panel: $m_{G_1} = 2$ {\rm TeV} and $\Lambda = 3$ {\rm TeV}.
The red dotted line represent the $\langle \sigma \, {\rm v} \rangle_{th} \equiv 2.2 \cdot 10^{-26} cm^3/s$. 
The blue dashed, orange dot-dashed and solid green lines represent the scalar, fermion and vector boson DM cases, respectively. 
In all plots the radion mass has been kept fixed to $m_r = 500$ {\rm GeV}
}
\label{fig:ejemplos_con_radion}
\end{figure*}

In Fig. \ref{fig:ejemplos_con_radion} we present the total thermally-averaged cross-section  $\langle \sigma \, {\rm v} \rangle_{th}$ as a function of the DM mass, 
for four different points in the parameter space: $(m_{G_1}, \Lambda) = (1,500)$ TeV (upper left panel); $(m_{G_1}, \Lambda) = (1,100)$ TeV (upper right panel);  
$(m_{G_1}, \Lambda) = (2,10)$ TeV (lower left panel); $(m_{G_1}, \Lambda) = (2,3)$ TeV (upper left panel). In all cases, the radion mass has been kept fixed to $m_r = 500$ GeV.
(notice that the actual value of the radion mass has no real impact onto the DM total annihilation cross-section, though).
In all panels we represent scalar, fermion and vector DM particles by dashed (blue), dot-dased (orange) and solid (green) lines, respectively. 
As in Fig.~\ref{fig:contribuciones}, the horizontal (red) dashed line and the red-shaded area represent the freeze-out thermal cross-section  $\langle \sigma_{\rm FO} \, {\rm v} \rangle$
and the region for which the effective field theory is not valid. 

We can see some generic features: 
(1) for vector boson DM, virtual KK-graviton exchange always dominates the cross-section;
(2) for scalar DM, the freeze-out cross-section is achieved only after the opening of the direct KK-graviton production channel;
(3) fermion DM has a much softer dependence on $m_{\rm DM}$ than scalar and vector boson DM (as it was already discussed
in Ref.~\cite{Folgado:2019gie});
 (4) the lower (the higher) $\Lambda$, the lower (the higher)  the DM mass for which the freeze-out cross-section is achieved.

\begin{figure*}[htbp]
\centering
\includegraphics[width=140mm]{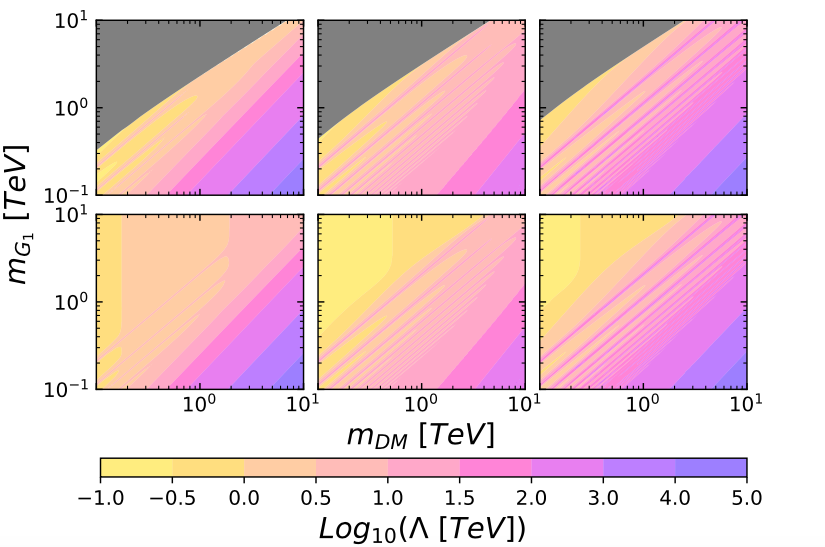}
\caption{From left to right we present the values of $\Lambda$  for which the observed DM relic abundance is obtained in the $(m_{DM}, m_{G_1})$ plane for scalar, fermion and vector boson DM particles.
Upper panels: the extra-dimension length is unstabilized; 
Lower panel: the extra-dimension length is stabilized using the Goldberger-Wise mechanism, with a radion mass $m_r = 100$ GeV. 
The required $\Lambda$ ranges from $10^{-1}$ to $10^5$ TeV, as shown by the color legend.}
\label{fig:lambda_scalars}
\end{figure*}

In order to understand the dependence of the DM annihilation cross-section on the three free paramers of the model $m_{\rm DM}, m_{G_1}$ and $\Lambda$,  
we show in Fig.~\ref{fig:lambda_scalars} the region of the $(m_{\rm DM},m_{G_1})$ plane for which $\langle \sigma_{\rm FO} \, {\rm v} \rangle$ is achievable, drawing the corresponding value of $\Lambda$ for which $\langle \sigma \, {\rm v} \rangle_{th} = \langle \sigma_{\rm FO} \, {\rm v} \rangle$. 
The upper panels represent our results in the case of an unstabilized extra-dimension, {\em i.e.} in the absence of the radion. 
On the other hand, in the lower panels we have included a radion accordingly to the Goldberger-Wise stabilization mechanism.  
Both in the upper and lower cases,  from left to right the three panels show the scalar, fermion and vector boson cases, respectively.
The main difference between the unstabilized and stabilized cases is the gray region in the upper left corner present for DM of any spin. This region represents the portion of
the parameter space for which the observed DM relic abundance cannot be achieved. We can see that, when no radion is present in the physical spectrum, 
the region at low DM mass and large $m_{G_1}$ is not able to reproduce $\langle \sigma_{\rm FO} \, {\rm v} \rangle$ for any value of $\Lambda$. On the other hand, when a radion 
is included, this region becomes accessible as the direct radion production channel $\sigma_{rr}$ opens for relatively low values of the radion mass, $m_{\rm DM} \geq m_r$.
Apart from this difference, the two rows are rather similar. The typical range of $\Lambda$ for which achieving $\langle \sigma_{\rm FO} \, {\rm v} \rangle$ is possible 
is $\Lambda \in [10^{-1},10^5]$ TeV. A periodic pattern in $\Lambda$ can be clearly seen for low $m_{\rm DM}$ for any spin of the DM particle, a consequence
of the fact that for these values of $m_{\rm DM}$ the freeze-out cross-section is achieved through the virtual KK-graviton exchange diagram (see Fig.~\ref{fig:contribuciones}).
We can also see that the scalar and vector boson cases are extremely similar for $m_{\rm DM} \geq 1$ TeV  (as it can also be seen in Fig.~\ref{fig:ejemplos_con_radion}, 
whenever  $\langle \sigma_{\rm FO} \, {\rm v} \rangle$ is achieved through direct KK-gravitons production). On the other hand, the range of $\Lambda$ for which 
the freeze-out cross-section is achievable in the fermion DM case is smaller, $\Lambda \in [10^{-1},10^3]$ TeV, as a consequence of the milder $m_{\rm DM}$ dependence
of the fermion DM annihilation cross-section. This points out that the fermion DM case will be more easily falsified by resonant searches at the LHC Run-III 
and its high-luminosity upgrade, the HL-LHC.

%%%%%%%%%%%%%%%%%%%%%%%%%%%%%%%%%%%%%%%%%%%%%%%%%%%%%%%%%%%%%%%%%%%%%%%%%%%
\section{Parameter space analysis}
\label{sec:results}
%%%%%%%%%%%%%%%%%%%%%%%%%%%%%%%%%%%%%%%%%%%%%%%%%%%%%%%%%%%%%%%%%%%%%%%%%%%

In this Section we search the different regions of the parameter space $(m_{DM}, m_{G_1}, \Lambda)$ for which is possible 
to achieve the correct relic abundance, $\langle \sigma \, {\rm v} \rangle_{th} = \langle \sigma_{\rm FO} \, {\rm v} \rangle$. We will first review briefly present experimental
bounds on the mass of the first KK-graviton and the effective gravitational scale $\Lambda$ and remind theoretical unitarity bounds on $m_{\rm DM}$. 
Eventually, in Fig.~\ref{fig:finalresults_conj} we show the region of the $(m_{\rm DM}, m_{G_1})$ plane for which the observed DM relic abundance is achieved
for scalar, fermion and vector boson DM, extending our previous results of Ref.~\cite{Folgado:2019sgz}.
%the different values of $\Lambda$ to obtain current DM yield, we founded that to each point of the parameter space it is possible to obtain a value of $\Lambda$ to obtain the correct %abundance. The idea in this section it is found the different theoretical and experimental bounds over the $(m_{G_1}, m_{DM}$ assuming that we always have the $\Lambda$ value 
%to reproduce the current quantity of DM.

\subsection{Experimental Bounds}

There are two kinds of experimental bounds to be imposed in the model parameter space: resonance searches at the LHC; and Direct and Indirect Dark Matter searches. We will review both kinds
of bounds in Sects.~\ref{sec:lhc}, \ref{sec:ddmd} and \ref{sec:idmd}.

\subsubsection{LHC bounds}
\label{sec:lhc}

The strongest constraints come from resonant searches at LHC Run II at $\sqrt{s} = 13 \, \text{TeV}$. In the RS model, two kinds of particles can be resonantly produced at the LHC: 
the radion and the KK-graviton tower. 
Out of the latter, bounds are usually imposed over the first KK-graviton mode, $G_1$, as in the absence of a signal we can only conclude that the mass of the corresponding resonance is larger
than the maximum avaiable energy to produce it. In the case a positive signal were to be found in the LHC Run III or at the HL-LHC, we should clearly look for more, heavier, resonances and check 
if the spacing between them is compatible with the values of $m_{G_n}$ expected in the model. 

In order to estimate the impact of the LHC Run II, it is necessary to analyse the production cross-section of these two kind of particles. The bound is over the production of bulk particles and 
it is independent of the DM mass and spin. The analysis realised in Ref.~\cite{Folgado:2019sgz}, therefore, is totally valid and it can be used in the three cases of scalar, fermion and vector boson DM particles. 
The conclusion of the study of the production was that the bounds on the resonant production of the radion are much weaker than those corresponding to KK-graviton production. 
Indeed, the $\bar{q} \, q \, r$ vertex is proportional to the corresponding quark mass and, then, resonant radion production is dominated by gluon-fusion at the considered energy. 
However, the interaction between gluons/photons and the radion arises through quarks and W boson loops via the trace anomaly \cite{Blum:2014jca}. Eventually, detection of resonant particles
at the LHC occurs dominantly in two possible ways,  $X \rightarrow \gamma \gamma$ and $X \rightarrow l\,  l$. However, radion decay to $\gamma \gamma$ and $l \, l$ is much smaller than the
corresponding decay of a KK-graviton. As a consequence, the overall bounds over $m_r$ are weaker than those over $m_{G_1}$, as anticipated above. Bounds over $m_{G_1}$ and $\Lambda$
from Refs.~\cite{Aaboud:2017yyg,ATLAS:2017wce,CMS:2018thv} are given in Fig.~7 of Ref.~\cite{Folgado:2019sgz}.

\subsubsection{Direct Dark Matter Detection}
\label{sec:ddmd}

Another possible source of experimental bounds is given by the DM searches at Direct Detection (DD) experiments.  These experiments are able to constrain the scattering cross-section between the DM particles and the nucleons of the experimental targets, with the cross-section parametrized by different operators.  In general,  the strongest bounds come from spin-independent terms in the DM-nucleon cross-section. 

In the present model we have two gravitational mediators: radion and KK-gravitons. The dominant contribution for the three DM spin cases considered here is always given by KK-gravitons. The reason is that the DM-quarks cross-section mediated by radions is suppressed by powers of the quark mass, and the interaction with gluons is generated via the trace anomaly. On the other hand, the interaction mediated by KK-gravitons do not suffer from any suppression. Therefore, the radion contribution to DD searches can safely be neglected in the analysis. 

In order to obtain the spin-independent DD cross-section it is sufficient to study the nuclear matrix elements for spin-2 mediators.  In principle, the functional form of the relevant operators depends on the spin of the DM. 
However, taking a zero momentum transfer for the DM-nucleon scattering,  the spin-independent cross-section only depends on the DM mass and it is given by  \cite{Carrillo-Monteverde:2018phy}:
\be
\sigma_{{\rm DM}-p}^{\rm SI} = \left [ \frac{m_p \, m_{\text{DM}} }{A \pi (m_{\text{DM}} + m_p)} \right ]^2\left [ Af_p + (A-Z)f_n  \right ]^2 \, ,  
\label{eq:DD}
\ee
where $m_p$ is the proton mass, $f_p$ and $f_n$ are the nucleon form factors and, eventually,  $Z$ and $A$ are the number of protons and the atomic number,  respectively.  We stress again that, for non-relativistic DM,  this expression is valid for the three DM cases. The nucleon form factors when KK-gravitons only couple to quarks can be found in Ref. \cite{Carrillo-Monteverde:2018phy}. In our case, all SM particles are confined to the IR-brane and the gravity mediators also couple to the gluons. However, quark and gluon contributions to the nucleon form factors are of the same order, and we do not expect huge differences  in the DD excluded area when the DM-gluon interaction is included.
Although it could change by ${\cal O}(1)$ factors, this means that for the cases considered here, the DD exclusion bounds are always  far looser than the LHC limits from resonance 
searches.

The strongest bounds from DD Dark Matter searches are found at XENON1T, which uses as target mass $^{129}$Xe, ($Z = 54$ and $A-Z = 75$). In order to compute the possible bounds over the three cases studied in the present work we use the exclusion curve of XENON1T \cite{Aprile:2017iyp} to set constraints in the $(m_{DM}, m_{G_1}, \Lambda)$ parameter space.

\subsubsection{Indirect Dark Matter Detection}
\label{sec:idmd}
Regarding DM indirect searches, there are several astrophysical experiments analysing different signals. 
The Fermi-LAT collaboration, for example, studied the $\gamma$-ray flux reaching Earth  from Dwarf spheroidal galaxies \cite{Fermi-LAT:2016uux}  
and the galactic center \cite{TheFermi-LAT:2015kwa,TheFermi-LAT:2017vmf}, while AMS-02 has reported data about the positrons \cite{PhysRevLett.113.221102} 
and anti-protons \cite{PhysRevLett.117.091103} arriving at Earth from the center of the galaxy. These results are relevant for DM  models that generate a continuum spectrum 
of different SM particles, such as the RS scenario we are considering. For the scalar and fermion DM cases there is a $d$- and a $p$-wave suppression, respectively, in the annihilation  into SM particles. For these two cases, only DM annihilation into KK-gravitons and radions may lead to observable signals. On the other hand, in the vector boson DM case  
all DM annihilation channels are $s$-wave and, therefore, it is the most interesting for this class of experiments. In principle, current experimental data for indirect detection of DM allows to constrain DM masses only below $\sim$ 100 GeV, since  for DM particles with mass above $\sim$ 1 TeV (as needed in our scenario to obtain the correct relic abundance)   the limits on the cross-section are well above 
the required value $\langle \sigma_{\rm FO} \, {\rm v} \rangle$. 

However, there is a caveat: in the region where the DM mass is much bigger than  the mediator mass $m_\phi$ (either the radion or the KK-graviton) and the DM coupling is large (but still allowed by unitarity and validity of the 4D effective field  theory, as discussed below), there could be a sizable Sommerfeld  enhancement of the DM annihilation cross section in the present highly non-relativistic regime. 
For instance, according to Ref.~\cite{Slatyer:2011kg}, for DM relative velocities $ {\rm v}$ smaller than the ratio of the mediator mass to the dark matter mass, 
${\rm v} < m_{\phi}/m_{\rm DM} $,  the Sommerfeld enhancement saturates due to the finite range of the force approaching the approximate value  
$g_D^2  \, m_{\rm DM}/\left [ m_{\phi}(1-\cos\theta) \right ]$, where $ g_D$ is the coupling between the DM and the mediator (given in our case by 
$m_{\rm DM}/\Lambda$),  and  $\theta \sim 2\sqrt{6} \sqrt{\alpha_D m_{\rm DM}/m_\phi}$, with $\alpha_D = g_D^2/(4\pi)$.
 Although the calculation in  \cite{Slatyer:2011kg} was done for fermionic DM and a scalar mediator, we shall assume that there would be a similar saturated enhancement in the case of KK-gravitons for all DM spins, in order to estimate the DM annihilation cross section today, as relevant for ID experiments.
From Fig.~\ref{fig:lambda_scalars}, we can see that the value of $\Lambda$ required to obtain the correct DM relic abundance in the relevant  allowed region where $m_{\rm DM}/m_{G_1} \gg 1$ 
(corresponding to $m_{\rm DM} \in [1,10]$ TeV and $m_{G_1} < 1$ TeV) is $\Lambda \sim 10^4$ TeV. Therefore, the angle $\theta$ is very small and we find that  the Sommerfeld enhancement factor is 
${\cal O}(1)$, i.e., no enhancement occurs.  In the case of DM annihilation  into radions, although the enhancement could be larger in a small region still allowed by LHC where $\Lambda \sim 1$ TeV 
(as shown in  Fig.~\ref{fig:finalresults_conj}),  the corresponding tree-level cross-section is always well below $\langle \sigma_{\rm FO} {\rm v} \rangle$ (see Fig.~\ref{fig:contribuciones}), so we do not expect any observable signal either.
 
 Thus,  we conclude that current  indirect searches have no impact on the viable parameter space for our scenario. Notice, however, that this could be tested in the next generation of ground-based observatory for  
 $\gamma$-rays, CTA \cite{CTAConsortium:2018tzg}.

\subsection{Theoretical limits}
\label{sec:unitarityandOPE}

Besides the experimental limits, there are two relevant theoretical assumptions to be fulfilled in order to ensure the validity of the approach used in this paper.
First, we have been performing a tree-level computation of the DM annihilation cross-sections, only. We must, therefore, worry about unitarity issues.
In particular, the t-channel annihilation cross-section into a pair of KK-gravitons, $\sigma_{GG}$, diverges as $m_{\rm DM}^8/(m_{G_n}^4 m_{G_m}^4)$ for scalar and vectorial DM particles and $m_{\rm DM}^4/(m_{G_n}^2 m_{G_m}^2)$ for spin $1/2$ particles in the non-relativistic limit $s \simeq m_{\rm DM}^2$. It is, therefore, mandatory to check that the effective theory is still unitary. 
We will take as unitarity bound that $ \sigma < 1/s \simeq 1/m_{\rm DM}^2$. This bound is shown in Fig.~\ref{fig:finalresults_conj} as a green-meshed area. 

Second, we should concern about the consistency of the effective theory framework. In a Randall-Sundrum framework, the effective scale of the theory is represented by $\Lambda$.
At energies much above this scale, KK-gravitons become strongly-coupled and the theory inherits the intrinsic non-renormalizability of the Einstein action, independently on 
the number of space-time dimensions. In this region, therefore, the effective field theory approach is no longer valid. We will force, then, $m_{G_1}$ to be less than $\Lambda$
in order to trust our results. As we are including the first KK-gravitons in the low-energy spectrum, they should be lighter than the effective field theory scale to be dynamical 
degrees of freedom of the theory. Notice that, in the allowed region, also the relation $m_{\rm DM} \leq \Lambda$ is automatically fulfilled.

\subsection{Results}

\begin{figure*}[htbp]
\centering
\includegraphics[width=140mm]{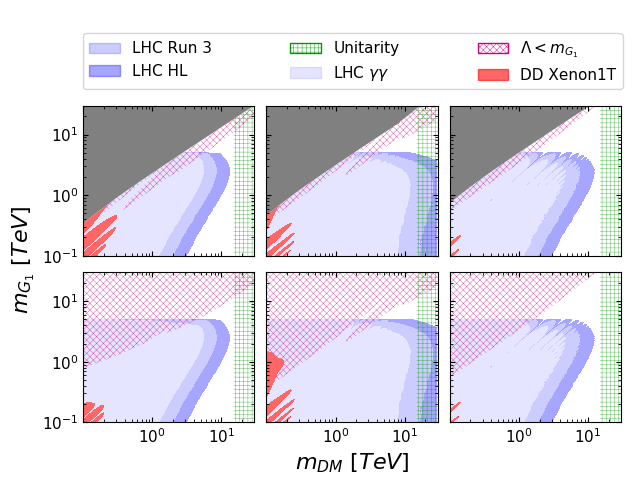}
\caption{\it Region of the $(m_{\rm DM},m_{G_1})$ plane for which $\langle \sigma \, {\rm v} \rangle_{th} = \langle \sigma_{FO} v \rangle $. Upper panels represent our results in the unstabilized case, {\em i.e.} 
when no radion is considered; lower panels depict the stabilized case, where the size of the extra-dimension is fixed by the Goldberger-Wise mechanism and a (light) radion is added to the 
spectrum. The radion mass in this case is $m_r = 100$ GeV. From left to right we present our results for scalar, fermion and vector boson DM particles. In all panels, the white (grey-shaded) area represents the 
region of the parameter space for which it is possible (impossible) to achieve the correct relic abundance. Over these regions, we have superimposed theoretical and experimental bounds. 
In particular, the pink-meshed area is the region for which the low-energy Randall-Sundrum effective theory is untrustable as $m_{G_1} < \Lambda$;  the vertical green-meshed area on the right 
of all panels is the region where the unitarity constraint is not fulfilled, $m_{DM} \, > \, 1/\sqrt{\sigma_{FO}}$; the red-shaded area is the region of the parameter space excluded by Direct Dark Matter Detection
searches; eventually, the three blue-shaded areas represent the region of the parameter space excluded by resonance searches at the LHC Run II with 36 fb$^{-1}$ (light blue) 
and foreseeably excluded by the LHC Run III with 300 fb$^{-1}$ (blue) and the HL-LHC with 3000 fb$^{-1}$ (dark blue). }
\label{fig:finalresults_conj}
\end{figure*}

We present our final results in the ($m_{\rm DM},m_{G_1}$) plane in Fig.~\ref{fig:finalresults_conj}, where the different panels represent the region of the parameter space
($m_{\rm DM},m_{G_1},\Lambda$)  for which the DM annihilation cross-section can  achieve the freeze-out value. From left to right, we show results for scalar, fermion
and vector boson Dark Matter, respectively. On the other hand, the difference between upper and lower plots stands in that in the upper plots the size of the extra-dimension is unstabilized, 
whereas in the lower ones we add the radion to the spectrum and implement the Goldberger-Wise mechanism to stabilize $r_c$.

In each of the panels, we depict by a white area the allowed region: this means that for each pair of values in the ($m_{\rm DM},m_{G_1}$) plane, it exists a specific value of $\Lambda$
for which $\langle \sigma \, {\rm v} \rangle_{th} = \langle \sigma_{\rm FO} \, {\rm v} \rangle$. The grey-shaded area, on the other hand, represent the region for which, for a particular choice in the 
($m_{\rm DM},m_{G_1}$) plane, no value of $\Lambda$ fulfills the freeze-out condition. We can see that a grey-shaded area exists in all of the three upper plots. This means that, 
in the absence of the radion, it always exists a region of the parameter space for which it is impossible to achieve $\langle \sigma_{\rm FO}\, {\rm v} \rangle$, independently of the spin 
of the Dark Matter particle. On the other hand, in all of the three lower panels the grey-shaded region is absent: it is always possible to reach  $\langle \sigma_{\rm FO} \, {\rm v} \rangle$
in the presence of a radion. This happens as the radion mass is not fixed: by choosing a conveniently light radion mass, the direct radion production channel $\sigma_{rr}$ 
gives an extra component to the total cross-section such that the observed relic abundance can be achieved. In all of the lower panels, we fix the radion mass to $m_r = 100$ GeV. 
Notice that bounds on the radion are much weaker than those on the first KK-graviton, as it was explained in Sect.~\ref{sec:lhc}.

On top of the allowed or disallowed regions, we draw the experimental bounds from Sects.~\ref{sec:lhc}, \ref{sec:ddmd} and \ref{sec:idmd}. The red-shaded area is the region of the 
parameter space incompatible with Direct Detection experiments. The peculiar periodic structure arises as for a fixed value of $m_{\rm DM}$ the correct relic abundance can be achieved
with multiple choices of the two other free parameters of the model, $m_{G_1}$ and $\Lambda$ (see Fig.~\ref{fig:ejemplos_con_radion} for a similar situation in a different plane). 
We see that this bound only constrains very low values of the Dark Matter mass, independently from the Dark Matter spin.
On the other hand, the light blue-shaded region is much more constraining: this corresponds to resonance searches at the LHC Run II, 
with a luminosity of ~36 fb$^{-1}$ at $\sqrt{s} = 13 \, \text{TeV}$ \cite{Aaboud:2017yyg,ATLAS:2017wce,CMS:2018thv}. 
In all cases, this bound is much stronger than those from DD and excludes Dark Matter masses below 1 TeV 
(or more, depending on the DM spin). The LHC bound saturates in $m_{G_1}$ around 5 TeV. Above this value, the LHC
is no longer able to push its bounds, independently from the luminosity, as the c.o.m. energy is not enough to produce the resonance.  This is not the case in the (horizontal) Dark Matter mass axis, 
as for this parameter increasing the LHC luminosity does make the bound stronger: this is depicted by increasingly darker blue-shaded areas, corresponding to the LHC Run III (with an expected
luminosity of 300 fb$^{-1}$) and to the foreseen LHC luminosity upgrade, the HL-LHC (with a goal luminosity of 3000 fb$^{-1}$).
Eventually, the green- and pink-meshed areas represent theoretical consistency and unitarity bounds from Sect.~\ref{sec:unitarityandOPE}. In particular, the pink-meshed area is 
the region of the parameter space for which the value of $\Lambda$ needed to achieve $\langle \sigma_{\rm FO} \, {\rm v} \rangle$ for a given point in the ($m_{\rm DM}, m_{G_1}$) plane
is lower than the first KK-graviton mass, $\Lambda < m_{G_1}$. In an OPE approach this condition is unviable, as we should integrate out particles heavier than the effective theory scale, 
in this case the whole tower of KK-gravitons. Notice that this constraint excludes most of the parameter space for which the observed relic abundance is achieved through direct radion production
(the region that opens in the upper left corner for DM of any spin in the lower panels, absent in the upper row).  The vertical green-meshed area in the rightmost side of each plot represents, 
on the other hand, the unitarity bound $m_{\rm DM} \leq 1/\sigma$.
This constraint puts an upper bound to the value of the Dark Matter mass for which the Randall-Sundrum model is able to explain the observed relic abundance within the freeze-out scenario.
Notice that, incidentally, in all of the allowed (white) region the Dark Matter mass is also smaller than the value of $\Lambda$ needed to achieve $\langle \sigma_{\rm FO} \, {\rm v} \rangle$, $m_{\rm DM} < \Lambda$. 

Once we have described what is common to all panels, we may now particularize to each DM spin case. The two leftmost plots correspond, as explained above, to the scalar DM case 
without (above) and with (below) a Goldberger-Wise radion. This case was already shown in Ref.~\cite{Folgado:2019sgz} and we get pretty similar results to those presented there
(the only difference being that in this case we have taken into account the DM DM $\to r \, G_n$ channel, previously overlooked). The region of the ($m_{\rm DM}, m_{G_1}$) plane
where it is possible to obtain the correct relic abundance and is not excluded by the theoretical and experimental bounds is dominated by direct graviton production. The virtual KK-graviton (and radion) exchange is always subdominant in this area. The difference between the unstabilized (above) and stabilized (below) cases is that in the latter it would be possible to reach the observed DM abundance 
for lower DM masses: this region, however, is excluded by the LHC Run II bounds for $m_{G_1} < 5$ TeV and by consistency of the effective theory for $m_{G_1} > 5$ TeV.

The two plots in the middle represent the spin 1/2 DM case. This case is the most constrained one between the three options studied here, as a consequence of the softer dependence of the 
cross-section on the DM mass (see Fig.~\ref{fig:ejemplos_con_radion}). The direct  KK-graviton production  channel in the fermion DM case diverges as $m_{DM}^4/m_{G_n}^2m_{G_m}^2$, and not 
as $m_{DM}^8/m_{G_n}^4m_{G_m}^4$, as it was the case for scalar and vector boson DM. The observed relic abundance is, therefore, reached later than for integer spin, closer to the region 
excluded by the unitarity limit, $m_{\rm DM} < 1/\sigma$. For spin 1/2 Dark Matter particles, the LHC bounds are extremely effective for $m_{G_1} < 5$ TeV, excluding all of the allowed region after taking into
account the unitarity bound. Both in the upper and lower panels we can see that only a tiny triangular region survives, for which 
$m_{G_1} > 5$ TeV, $m_{\rm } \in [4,15]$ TeV and $\Lambda > m_{G_1}$. Notice that our results for the fermion case are those that differ the most with respect to the CW/LD case: in the latter, 
the requirement that $m_{G_1}$ and $m_{\rm DM}$ be smaller than the gravitational scale $M_5$ excludes all of the region above the diagonal $m_{G_1} > m_{\rm DM}$, and the combination with the LHC
bound implies that the only viable region is for large $m_{\rm DM}$ ($m_{\rm DM} > 5$ TeV) and low $m_{G_1}$ ($k < 300$ GeV), corresponding to $M_5 > 10$ TeV. On the other hand, we have seen
that for the RS case a viable region at large $m_{\rm DM}$ ($m_{\rm DM} > 5$ TeV) and large $m_{G_1}$ ($m_{G_1} >  5$ TeV) can be found, corresponding to relatively low values of $\Lambda$ 
($\Lambda < 10$ TeV), i.e. values that could alleviate the hierarchy problem.

Eventually, the vector boson DM case is depicted in the two rightmost panels. This is the only one for which the virtual KK-graviton and radion exchange channels have some effect in the phenomenology 
in the allowed region. The periodic pattern caused by the dominance of these channels in some part of the parameter space induces the peculiar wiggled behaviour in the upper right corner of the LHC experimental bounds. The surviving allowed (white) region is very similar to what we got in the scalar DM case, as the cross-section dependence on the DM mass is analytically the same.

%%%%%%%%%%%%%%%%%%%%%%%%%%%%%%%%%%%%%%%%%%%%%%%%%%%%%%%%%%%%%%%%%%%%%%%%%%%
\section{Conclusions}
\label{sec:con}
%%%%%%%%%%%%%%%%%%%%%%%%%%%%%%%%%%%%%%%%%%%%%%%%%%%%%%%%%%%%%%%%%%%%%%%%%%%

In this paper we have completed the analysis started in Ref.~\cite{Folgado:2019sgz}, checking the viability of the hypothesis that the observed Dark Matter relic abundance in the Universe
may be explained, within the context of the freeze-out mechanism, by gravitationally-interacting massive particles embedded in a Randall-Sundrum \cite{Randall:1999ee} extra-dimensional model. Whereas in Ref.~\cite{Folgado:2019sgz} 
we studied scalar DM particles, only, we have extended
here the analysis to spin 1/2 and spin 1 particles, showing that in all cases the observed relic abundance can be reproduced in the proposed framework.
This happens as the, otherwise exceedingly small, gravitational interaction is enhanced in extra-dimensional models either by the volume of the extra-dimension or by their curvature 
(being this latter option the one at work in our case). This paper put our study of the Randall-Sundrum extra-dimensional DM scenario on equal footing with an analogue search
that we presented in Ref.~\cite{Folgado:2019gie}, where we studied the same possibility in  the Clockwork/Linear Dilaton extra-dimensional model for DM particles of spin 0, 1/2 and 1.  
%We have found that a region of the parameter space for which achieving the observed DM relic abundance
%is compatible with present and future experimental and theoretical constraints can be found  in both models.

In both the RS and the CW/LD models two branes are considered, the so-called UV (or Planck) and IR (or TeV) branes. Standard Model matter is traditionally constrained to the IR-brane in both cases. We also 
choose to constrain the Dark Matter particle, whichever its spin, to the IR-brane.  In this particular scenario the interaction between two particles located in the IR-brane via gravity is proportional to 
$1/M_{\rm P}^2$ when the interaction occurs thanks to the Kaluza-Klein zero-mode ({\em i.e.} the standard graviton), whereas the interaction with higher Kaluza-Klein modes is suppressed only by two powers
of the effective scale $\Lambda$. Since $\Lambda$ can be as low as a few TeV (so as to solve the so-called {\em hierarchy problem}, the original motivation for the existence of extra-dimensions), a huge enhancement in the cross-section is possible with respect to standard linearized General Relativity. In addition to the KK-tower of gravitons, we also consider a radion field, added in such a way so as 
to stabilize the size of the extra-dimension taking advantage of the Goldberger-Wise mechanims. Other possibilties could be (and have been) considered, such as allowing for the Dark Matter to freely explore the bulk. However, we have found that also in our restrictive case the freeze-out mechanism is efficient enough to explain the observed DM relic abundance. 

We have then computed the different contributions to the thermally-averaged DM annihilation cross-section $\langle \sigma \, {\rm v} \rangle$ for each of the three DM particles studied here with 
spin 0, 1/2 and 1. The channels considered for the analysis are the virtual KK-gravitons and radion exchange and the direct production of two ``gravitational" modes (either two KK-gravitons, or
one KK-graviton and one radion, or two radions). As a consequence of the polarization of the spin-2 KK-gravitons, the dominant channel for any of the considered DM spins is the direct production
of two KK-gravitons, when the DM mass is larger than 1 TeV, approximately.
 In the scalar and vector cases the corresponding cross-section is enhanced at large DM masses by a term proportional to  $m_{DM}^8/(m_{G_n}^4 m_{G_m}^4)$.
 In contrast with the spin 0 and 1 cases, the cross-section for direct KK-gravitons production in the spin 1/2 case is enhanced by a softer factor,  $m_{DM}^4/(m_{G_n}^2 m_{G_m}^2)$. 
 As a consequence, the observed relic abundance for spin $1/2$ DM particles is achieved at larger values of the DM mass where, however, the unitarity bound on the DM mass takes over.
 
 We have scanned the three-dimensional parameter space of the model, ($m_{\rm DM}, m_{G_1}, \Lambda$), looking for the regions for which $\langle \sigma \, {\rm v} \rangle_{th} = \langle \sigma_{FO} v \rangle$
 whilst being compatible with present and foreseeable theoretical and experimental bounds. Our results were eventually shown in Fig.~\ref{fig:finalresults_conj}. We have found that the most
 relevant experimental constraint comes from LHC Run II resonance searches, whereas Direct and Indirect Dark Matter Detection experiments are mostly irrelevant for DM masses above 1 TeV.
 The theoretical requirements that $m_{\rm DM} < 1/\sigma$ and that $\Lambda$ be larger than $m_{\rm DM}, m_{G_1}$ constrain significantly the parameter space, also.

Our main result is that a significant portion of parameter space in the $(m_{DM} , m_{G_1} )$ plane is able to reproduce the current data about the DM relic abundance for any of the considered DM spins. Most part
of the allowed region is, however, excluded by theoretical and experimental bounds. This is particularly true in the case of spin 1/2 Dark Matter, for which only a tiny triangular region survives 
for  $m_{G_1} > 5$ TeV, $m_{\rm DM} \in [4,15]$ TeV and $\Lambda > m_{G_1}$ (but tipically {\em smaller than} or around 10 TeV). This region can only be explored by accelerators with more c.o.m. energy than the LHC. On the other hand, both for scalar and vector boson DM particles, the LHC and its upgrades cannot exclude a region with $m_{\rm DM} \in [4,15]$ TeV and $m_{G_1} < 10$ TeV. In this region, $\Lambda$ ranges from a few TeV to $10^4$ TeV, approximately. Notice that in most part of the allowed regions, for DM of any of the spins considered here, the hierarchy problem cannot be fully solved and a (softer) hierarchy is still present between $\Lambda$ and the electro-weak scale $\Lambda_{\rm EW}$. We have found that the presence or absence of the radion is mostly irrelevant and our results do not depend on it.

\section*{Acknowledgements} 

We thank   Hyun Min Lee, Roberto Ruiz de Austri  and Ver\'onica Sanz for illuminating discussions.
This work has been partially supported by the European Union projects H2020-MSCA-RISE-2015 and H2020-MSCA- ITN-2015//674896-ELUSIVES,
by the Spanish MINECO under grants  FPA2017-85985-P and  SEV-2014-0398, and by Generalitat Valenciana through the ``plan GenT" program (CIDEGENT/2018/019)
and grant PROMETEO/2019/083.

\appendix

\section{Feynman rules}
\label{app:feynman}
%%%%%%%%%%%%%%%%%%%%%%%%%%%%%%%%%%%%%%%%%%%%%%%%%%%%%%%%%%%%%%%%%%%%%%%%%%%

We remind in this Appendix the different Feynman rules corresponding to the couplings of DM particles and of SM particles with KK-gravitons and radion/KK-dilatons. In \cite{Folgado:2019sgz} we give the Feynman rules for the scalar case, in this appendix we show a complete description to any spin.

\subsection{Graviton Feynman rules}
\label{app:gravFR}

The vertex that involves one KK-graviton and two scalars $S$ of mass $m_S$ is given by:
\bea
\qquad
\raisebox{-15mm}{\includegraphics[keepaspectratio = true, scale = 1] {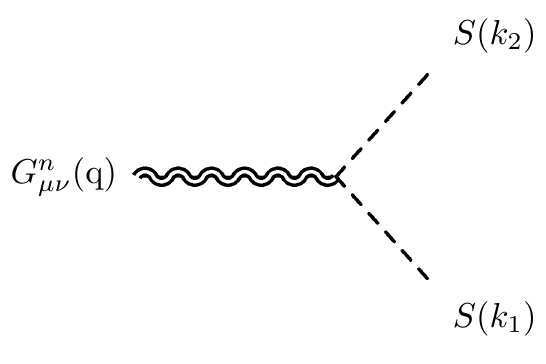}} \nonumber \\
\begin {aligned}
=-\frac{i}{\Lambda} \left ( m^2_S \eta_{\mu \nu} - C_{\mu \nu \rho \sigma} k_1^{\rho} k_2^{\sigma} \right ) \, ,
\end {aligned}
\eea
where
\be
C_{\mu \nu \alpha \beta} \equiv \eta_{\mu \alpha} \eta_{\nu \beta} + \eta_{\nu \alpha} \eta_{\mu \beta} - \eta_{\mu \nu} \eta_{\alpha \beta} \, .
\ee
This expression can be used for the coupling of both scalar DM and the SM Higgs boson to gravitons.

The vertex that involves one KK-graviton and two fermions $\psi$ of mass $m_\psi$ is given by: 
\bea
\qquad
\raisebox{-15mm}{\includegraphics[keepaspectratio = true, scale = 1] {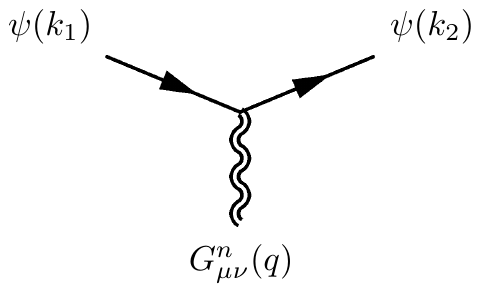}} \nonumber \\
\begin {aligned}
=& - \frac{i}{4\Lambda}
\left [ \gamma_{\mu} \left ( k_{2 \nu}+k_{1 \nu} \right ) + \gamma_{\nu} \left ( k_{2 \mu}+k_{1 \mu} \right ) \right. \\
& \left. - 2 \eta_{\mu \nu}\left ( \slashed{k_2}+\slashed{k_1}-2m_{\psi} \right )\right ] \, ,
\end {aligned}
\eea
and
\bea
\qquad
\raisebox{-15mm}{\includegraphics[keepaspectratio = true, scale = 1] {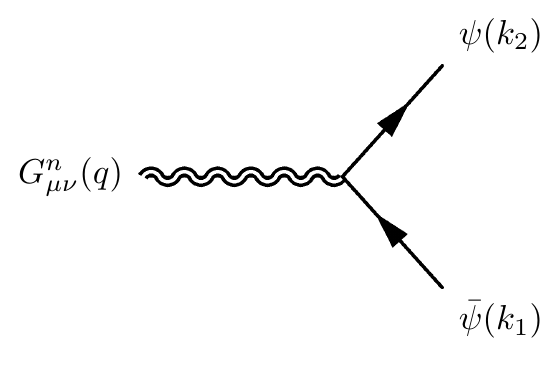}} \nonumber \\
\begin {aligned}
=& - \frac{i}{4\Lambda}\left [ \gamma_{\mu} \left (k_{2 \nu} - k_{1 \nu} \right ) 
+ \gamma_{\nu} \left (k_{2 \mu} - k_{1 \mu} \right ) \right. \\
&\left. - 2 \eta_{\mu \nu}\left ( \slashed{k_2} -\slashed{k_1} -2m_{\psi} \right )\right ] \, .
\end {aligned}
\eea

The interaction between two vector bosons $V$ of mass $m_V$ and one KK-graviton is given by:
\bea
\qquad
\raisebox{-15mm}{\includegraphics[keepaspectratio = true, scale = 1] {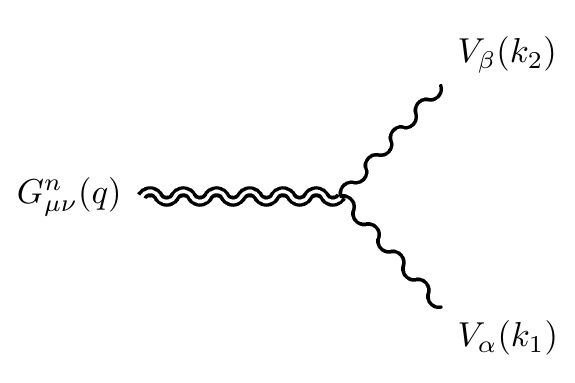}}\nonumber \\
\begin {aligned}
=-\frac{i}{\Lambda} \left ( m^2_V C_{\mu \nu \alpha \beta} + W_{\mu \nu \alpha \beta} \right ) \, ,
\end {aligned}
\eea
where
\be
W_{\mu \nu \alpha \beta} \equiv B_{\mu \nu \alpha \beta} + B_{\nu \mu \alpha \beta}
\ee
and
\bea
B_{\mu \nu \alpha \beta} &\equiv& \eta_{\alpha \beta}k_{1 \mu}k_{2 \nu} + \eta_{\mu \nu}(k_1 \cdot k_2 \eta_{\alpha \beta} - k_{1 \beta} k_{2 \nu}) \nonumber \\
&-& \eta_{\mu \beta} k_{1 \nu} k_{2 \alpha} + \frac{1}{2}\eta_{\mu \nu}(k_{1 \beta} k_{2 \alpha} - k_1 \cdot k_2 \eta_{\alpha \beta}) \, .
\eea

Eventually, the interaction between two particles ($S, \psi$ or $V_\mu$ depending on their spin) and two KK-gravitons (coming from a second order expansion 
of the metric $g_{\mu\nu}$ around the Minkowski metric $\eta_{\mu\nu}$) is given by:
\bea
\label{vertex:scalar_scalar_graviton_graviton}
\qquad
\raisebox{-15mm}{\includegraphics[keepaspectratio = true, scale = 1] {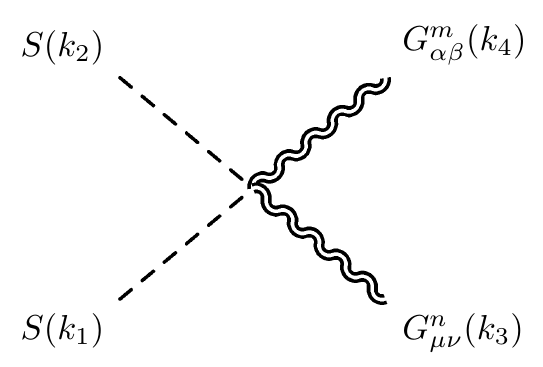}} \nonumber \\
\begin {aligned}
=& -\frac{i}{\Lambda^2} \eta_{\nu \beta} \left ( m^2_S \eta_{\mu \alpha} - C_{\mu \alpha \rho \sigma} k_1^{\rho} k_2^{\sigma} \right ) \, ,
\end {aligned}
\eea
\bea
\label{vertex:fermion_fermion_graviton_graviton}
\qquad
\raisebox{-15mm}{\includegraphics[keepaspectratio = true, scale = 1] {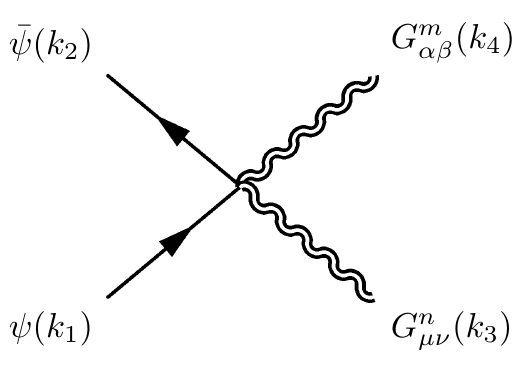}}\nonumber \\
\begin {aligned}
=& -\frac{i}{\Lambda^2} \eta_{\nu \beta} \left [ \gamma_{\mu} \left (k_{1 \alpha} - k_{2 \alpha} \right ) 
+ \gamma_{\alpha} \left (k_{1 \mu} - k_{2 \mu} \right ) \right. \\
&\left. - 2 \eta_{\mu \alpha}\left ( \slashed{k_1} -\slashed{k_2} -2m_\psi \right )\right ] \, ,
\end {aligned}
\eea
\bea
\label{vertex:vector_vector_graviton_graviton}
\qquad
\raisebox{-15mm}{\includegraphics[keepaspectratio = true, scale = 1] {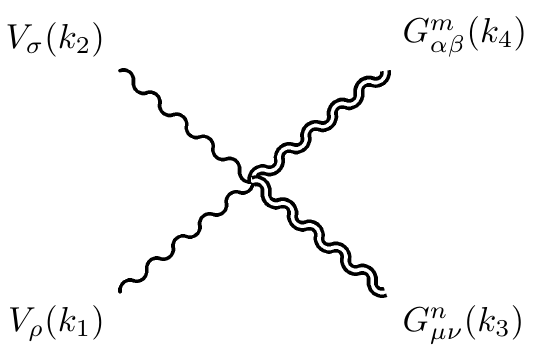}} \nonumber \\
\begin {aligned}
=& -\frac{i}{\Lambda^2} \eta_{\nu \beta} \left ( m^2_V C_{\mu \alpha \rho \sigma} + W_{\mu \alpha \rho \sigma} \right ) \, .
\end {aligned}
\eea

The Feynman rules for the $n=0$ KK-graviton can be obtained by the previous ones by replacing $\Lambda$ with $M_{\rm P}$. 
We do not give here the triple KK-graviton vertex, as it is irrelevant for the phenomenological applications of this paper.

\bibliography{bibliografia} 

\end{document}